\documentclass[superscriptaddress,aps,prd,nofootinbib,twocolumn,showpacs]{revtex4-1}

\usepackage{multirow}
\usepackage{amsmath,epsfig,ulem}
\usepackage{amsfonts}
\usepackage{amsmath}
\usepackage{amssymb}
\usepackage{graphicx}
\usepackage{hhline,color}
\usepackage{pifont}
\usepackage{lmodern}
\usepackage[utf8]{inputenc}

\newcommand{\unit}[1]{\ensuremath{\, \mathrm{#1}}}
\newcommand{\pr}[1]{\ensuremath{\left[#1\right]}}
\newcommand{\pc}[1]{\ensuremath{\left(#1\right)}}
\newcommand{\chav}[1]{\ensuremath{\left\{#1\right\}}}
\newcommand{\ev}[1]{\ensuremath{\left\langle #1\right\rangle}}

\def\beq{\begin{equation}}
\def\eeq{\end{equation}}
\def\beqa{\begin{eqnarray}}
\def\eeqa{\end{eqnarray}}
\def\ban{\begin{eqnarray*}}
\def\ean{\end{eqnarray*}}
\def\bi{\begin{itemize}}
\def\ei{\end{itemize}}

\newcommand{\Z}{\mathbb{Z}}

\begin{document}

\title{
Deconfinement, chiral symmetry restoration and thermodynamics of (2+1)--flavor 
hot QCD matter in an external magnetic field
}

\author{M\'arcio Ferreira}
\email{mferreira@teor.fis.uc.pt}
\affiliation{Centro de F\'{\i}sica Computacional, Department of Physics,
University of Coimbra, P-3004 - 516  Coimbra, Portugal}
\author{Pedro Costa}
\email{pcosta@teor.fis.uc.pt}
\affiliation{Centro de F\'{\i}sica Computacional, Department of Physics,
University of Coimbra, P-3004 - 516  Coimbra, Portugal}
\author{Constan\c ca Provid\^encia}
\email{cp@teor.fis.uc.pt}
\affiliation{Centro de F\'{\i}sica Computacional, Department of Physics,
University of Coimbra, P-3004 - 516  Coimbra, Portugal}

\date{\today}

\begin{abstract}
The entanglement extended Polyakov--Nambu--Jona-Lasinio model at zero chemical 
potential in the presence of an external magnetic field is studied. 
The effect of the entanglement parametrization is analyzed, in particular, on the
pseudocritical transition temperatures and on the thermodynamical properties of the model. 
The model predicts that the coincidence or not of both chiral and deconfinement transition
temperatures, in the presence of an external magnetic field, depends on the entanglement 
parametrization chosen.
\end{abstract}

\pacs{24.10.Jv, 11.10.-z, 25.75.Nq}

\maketitle

\section{Introduction}

Understanding matter under extremely intense magnetic fields is one of the 
most interesting topics in modern physics due to its relevance for studies involving compact
objects like magnetars \cite{duncan}, measurements in heavy ion collisions at very high
energies \cite{HIC,Kharzeev:2007jp} or the first phases of the universe \cite{cosmo}. 
The properties of the quark-gluon plasma (QGP) is a long-standing theoretical issue since the
discovery of the asymptotic freedom of QCD.
The structure of the QCD phase diagram in the presence of an external
magnetic field has been the subject of several studies
\cite{Klimenko:1991he,Ebert:2003yk,Ferrer:2005vd,Mizher:2010zb,Chatterjee:2011ry,Chernodub:2011mc}, 
in particular, at zero chemical potential $\mu=0$ (the $T-eB$ plane) (see
\cite{Bali:2011qj,Gatto:2012sp,Fraga:2012rr,D'Elia:2012tr} for a review). 

At zero chemical potential, almost all low-energy effective models, 
including the NJL-type models, as well as some lattice QCD (LQCD) calculations
\cite{D'Elia:2010nq,D'Elia:2011zu,Braguta:2010ej,Ilgenfritz:2012fw,Ilgenfritz:2013ara}, 
found an enhancement of the condensate due to the magnetic field (magnetic catalysis) 
independently of the temperature. 
However, a recent LQCD study \cite{Bali:2011qj,Bali:2012zg}, for $N_f=2+1$ flavors 
with physical quarks and pion masses, shows a different behavior in the transition 
temperature region, in particular, the suppression of the light condensates 
by the magnetic field, an effect known as inverse magnetic catalysis
\cite{Chao:2013qpa,Fukushima:2012kc,Kojo:2012js}. 
The reaction of the gluon sector to the presence of an external magnetic field 
should be incorporated into effective models in order to describe the inverse magnetic catalysis 
\cite{Bruckmann:2013oba}. One way to take it into account is to choose a magnetic field dependent
$T_0(eB)$ of the Polyakov potential \cite{Ferreira:2013tba}.

The Polyakov-loop extended Nambu--Jona-Lasinio model (PNJL) \cite{Fukushima:2003fw.Ratti:2005jh} 
has been generalized to include an effective four-quark vertex interaction depending 
on the Polyakov loop, the entanglement interaction \cite{Sakai:2010rp}.
This extension is known as the entanglement extended PNJL model (EPNJL). 
The entanglement interaction generates the correlation between the chiral restoration 
and deconfinement transition needed to be consistent with LQCD results at imaginary 
isospin quark-number chemical potential and real and imaginary chemical potentials.
An equation of state was constructed and the phase diagram in SU(2) were studied 
in \cite{Sakai:2011fa} using the EPNJL model. 
The theta-vacuum effects on the QCD phase diagram was studied in \cite{Sasaki:2011cj} 
using the SU(3) EPNJL model. The three-flavor phase diagram for zero and
imaginary quark-number chemical potential using EPNJL was performed in \cite{Sasaki:2011wu} 
with the entanglement interaction being parametrized in order to reproduce qualitatively the SU(3) 
LQCD results at zero and imaginary chemical potentials \cite{Aoki:2006we,deForcrand:2010he}.
In \cite{Gatto:2010pt} the phase diagram of QCD in an external magnetic field was studied 
using the SU(2) EPNJL model with and without 8-quark interaction \cite{8quarks}. 
Our aim is to extend the study of the $T-eB$ plane using the ($2+1$)-flavor (E)PNJL models 
including the 't Hooft determinant that reproduces $U_A(1)$ anomaly, responsible for the
mechanism of flavor mixing. In particular, we want to determine how the entanglement 
interaction is affected by the magnetic field and its consequences in the model predictions.

The effect of external magnetic fields on deconfinement and chiral pseudocritical
temperatures has been discussed in \cite{Gatto:2012sp,Fukushima:2010fe, Gatto:2010qs} 
using both the SU(2) PNJL and EPNJL models.
As in almost all other low-energy QCD models, these two models predict that 
the  critical temperature for chiral symmetry restoration increases with the 
increase of an external magnetic field strength. 
It was also shown that  within the EPNJL the splitting between the chiral 
and deconfinement transition temperatures is smaller than the splitting 
predicted by the PNJL model \cite{Gatto:2010pt}, and at $eB=19m_{\pi}^2$
it is not larger than $2\%$.
The phase diagram of 2+1 flavor PNJL model with charge asymmetry under an external 
magnetic field was also investigated in \cite{Costa:2013zca}, while the effects 
of an external magnetic field on the fluctuations of quark number, fluctuations and 
correlations of conserved charges was studied in \cite{Fu:2013ica}.

This paper is organized as follows: in Sec. II we present the model and the formalism 
starting with the deduction of the self-consistent equations. We also extract the equations 
of state and the thermodynamical quantities that will be studied.
In Sec. III the effect of different parametrizations of the entanglement interaction
on the transition temperatures is studied. The effect of the parametrization
of the entanglement interaction at zero chemical potential on the thermodynamical
quantities is carried out in Sec. IV, and some conclusions are drawn in the last
section. 

\section{Model and Formalism}
\label{sec:model}

\subsection{Model Lagrangian and gap equations}

We describe three flavor ($N_c=3$) quark matter subject to strong magnetic fields 
within the 2+1 EPNJL model.
The PNJL Lagrangian with explicit chiral symmetry breaking, where the quarks couple 
to a (spatially constant) temporal background gauge field, represented in terms of the
Polyakov loop, and in the presence of an external magnetic field is given by 
\cite{Fukushima:2003fw.Ratti:2005jh}:
\begin{eqnarray}
{\cal L} &=& {\bar{q}} \left[i\gamma_\mu D^{\mu}-
	{\hat m}_c \right ] q ~+~ {\cal L}_{sym}~+~{\cal L}_{det} \nonumber\\
&+& \mathcal{U}\left(\Phi,\bar\Phi;T\right) - \frac{1}{4}F_{\mu \nu}F^{\mu \nu},
	\label{Pnjl}
\end{eqnarray}
where the quark sector is described by the  SU(3) version of the Nambu--Jona-Lasinio model which
includes scalar-pseudoscalar and the 't Hooft six fermion interactions that
models the axial $U_A(1)$ symmetry breaking  \cite{Hatsuda:1994pi.Klevansky:1992qe},
with ${\cal L}_{sym}$ and ${\cal L}_{det}$  given by \cite{Buballa:2003qv},
\begin{eqnarray}
	{\cal L}_{sym}= \frac{G}{2} \sum_{a=0}^8 \left [({\bar q} \lambda_ a q)^2 + 
	({\bar q} i\gamma_5 \lambda_a q)^2 \right ] ,
\end{eqnarray}
\begin{eqnarray}
	{\cal L}_{det}=-K\left\{{\rm det} \left [{\bar q}(1+\gamma_5)q \right] + 
	{\rm det}\left [{\bar q}(1-\gamma_5)q\right] \right \}
\end{eqnarray}
where $q = (u,d,s)^T$ represents a quark field with three flavors, 
${\hat m}_c= {\rm diag}_f (m_u,m_d,m_s)$ is the corresponding (current) mass matrix,
$\lambda_0=\sqrt{2/3}I$  where $I$ is the unit matrix in the three flavor space, 
and $0<\lambda_a\le 8$ denote the Gell-Mann matrices.
The coupling between the (electro)magnetic field $B$ and quarks, and between the 
effective gluon field and quarks is implemented  {\it via} the covariant derivative 
$D^{\mu}=\partial^\mu - i q_f A_{EM}^{\mu}-i A^\mu$ where $q_f$ represents the 
quark electric charge ($q_d = q_s = -q_u/2 = -e/3$),  $A^{EM}_\mu$ and 
$F_{\mu \nu }=\partial_{\mu }A^{EM}_{\nu }-\partial _{\nu }A^{EM}_{\mu }$ 
are used to account for the external magnetic field and 
$A^\mu(x) = g_{strong} {\cal A}^\mu_a(x)\frac{\lambda_a}{2}$ where
${\cal A}^\mu_a$ is the SU$_c(3)$ gauge field.
We consider a  static and constant magnetic field in the $z$ direction, 
$A^{EM}_\mu=\delta_{\mu 2} x_1 B$.
In the Polyakov gauge and at finite temperature the spatial components of the 
gluon field are neglected: 
$A^\mu = \delta^{\mu}_{0}A^0 = - i \delta^{\mu}_{4}A^4$. 
The trace of the Polyakov line defined by
$ \Phi = \frac 1 {N_c} {\langle\langle \mathcal{P}\exp i\int_{0}^{\beta}d\tau\,
A_4\left(\vec{x},\tau\right)\ \rangle\rangle}_\beta$
is the Polyakov loop which is the {\it exact} order parameter of the $\Z_3$ 
symmetric/broken phase transition in pure gauge.

To describe the pure gauge sector an effective potential $\mathcal{U}\left(\Phi,\bar\Phi;T\right)$
is chosen in order to reproduce the results obtained in lattice calculations \cite{Roessner:2006xn},
\begin{eqnarray}
	& &\frac{\mathcal{U}\left(\Phi,\bar\Phi;T\right)}{T^4}
	= -\frac{a\left(T\right)}{2}\bar\Phi \Phi \nonumber\\
	& &
	+\, b(T)\mbox{ln}\left[1-6\bar\Phi \Phi+4(\bar\Phi^3+ \Phi^3)-3(\bar\Phi \Phi)^2\right],
	\label{Ueff}
\end{eqnarray}
where $a\left(T\right)=a_0+a_1\left(\frac{T_0}{T}\right)+a_2\left(\frac{T_0}{T}\right)^2$, 
$b(T)=b_3\left(\frac{T_0}{T}\right)^3$.
The standard choice of the parameters for the effective potential $\mathcal{U}$ is
$a_0 = 3.51$, $a_1 = -2.47$, $a_2 = 15.2$, and $b_3 = -1.75$.

As is well known, the effective potential exhibits the feature of a phase transition from 
color confinement ($T<T_0$, the minimum of the effective potential being at $\Phi=0$) to color
deconfinement ($T>T_0$, the minimum of the effective potential occurring at $\Phi \neq 0$).

We know that the parameter $T_0$ of the Polyakov potential defines the onset of deconfinement
and is normally fixed to $270$ MeV according to the critical temperature for the deconfinement
in pure gauge lattice findings (in the absence of dynamical fermions) \cite{Kaczmarek:2002mc}. 
When quarks are added to the system, quark backreactions can be taken into account,
thus a decrease in $T_0$ to $210\unit{MeV}$ is needed to obtain the deconfinement
pseudocritical temperature given by LQCD, within the PNJL model. 
Therefore, the value of $T_0$ is fixed in order to reproduce LQCD results 
($\sim$ 170 MeV \cite{Aoki:2009sc}).

The coupling constant $G$ in ${\cal L}_{sym}$ denotes the scalar-type four-quark interaction 
of the NJL sector. To obtain the EPNJL model, we substitute $G$ by $G(\Phi, \bar{\Phi})$, which 
depends on the Polyakov loop. 
As already mentioned, this effective vertex generates entanglement interactions between the 
Polyakov loop and the chiral condensate \cite{Sakai:2010rp}.   
The functional form of $G(\Phi, \bar{\Phi})$ was introduced in \cite{Sakai:2010rp} and reads,
\begin{eqnarray}
G(\Phi,
\bar{\Phi})=G\left[1-\alpha_1\Phi\bar{\Phi}-\alpha_2(\Phi^3+\bar{\Phi}^3)\right].
\label{G}
\end{eqnarray}
Also, for the EPNJL model we use $T_0 = 210$ MeV.

Once the model is not renormalizable, we use as a regularization scheme,
a sharp cutoff, $\Lambda$, in three-momentum space, only for the divergent 
ultraviolet integrals. 
The parameters of the model, $\Lambda$, the coupling constants $G$ and $K$,
and the current quark masses $m_u^0$ and $m_s^0$ are determined  by fitting
$f_\pi$, $m_\pi$ , $m_K$ and $m_{\eta'}$ to their empirical values. 
We consider $\Lambda = 602.3$, MeV, $m_u= m_d=5.5$, MeV,
$m_s=140.7$ MeV, $G \Lambda^2= 3.67$ and $K \Lambda^5=12.36$
as in \cite{Rehberg:1995kh}.
The thermodynamical potential for the three-flavor quark sector $\Omega$ is written as
\begin{align}
\Omega(T,\mu)&=G(\Phi,\bar{\Phi})\sum_{i=u,d,s}\ev{\bar{q}_iq_i}^2
+4K\ev{\bar{q}_uq_u}\ev{\bar{q}_dq_d}\ev{\bar{q}_sq_s} \nonumber \\
+&{\cal U}(\Phi,\bar{\Phi},T)+\sum_{i=u,d,s}\pc{\Omega_{\text{vac}}^i+\Omega_{\text{med}}^i
+\Omega_{\text{mag}}^i}
\end{align}
where the flavor contributions from vacuum $\Omega^{\text{vac}}_i$, medium $\Omega^{\text{med}}_i$, 
and magnetic field $\Omega^{\text{mag}}_i$ \cite{Menezes:2008qt.Menezes:2009uc} are given by
\begin{align}
 \Omega_{\text{vac}}^i&=-6\int_{\Lambda}\frac{d^3p_i}{(2\pi)^3}E_i\\
\Omega_{\text{med}}^i&=-T\frac{|q_iB|}{2\pi}\sum_{n=0}\alpha_n\int_{-\infty}^{+\infty}\frac{dp_z^i}{2\pi}\pc{Z_{\Phi}^+(E_i)+Z_{\Phi}^-(E_i)}\\
 \Omega_{\text{mag}}^i&=-\frac{3(|q_i|B)^2}{2\pi^2}\pr{\zeta^{'}(-1, x_i)-\frac{1}{2}(x_i^2-x_i)\ln x_i+\frac{x_i^2}{4}} 
\end{align}
where $E_i=\sqrt{(p_z^i)^2+M_i^2+2|q_i|Bk}$ , $\alpha_0=1$ and $\alpha_{k>0}=2$,
$x_i=M_i^2/(2|q_i|B)$, and $\zeta^{'}(-1, x_i)=d\zeta(z, x_i)/dz|_{z=-1}$, 
where $\zeta(z, x_i)$ is the Riemann-Hurwitz zeta function. 
The distribution functions $Z_{\Phi}^+$ and $Z_{\Phi}^-$ read
\begin{align}
 Z_{\Phi}^+&=\ln\chav{1+3\bar{\Phi}e^{-\beta (E_i-\mu)}+3\Phi e^{-2\beta (E_i-\mu)}+e^{-3\beta (E_i-\mu)}}\\
  Z_{\Phi}^-&=\ln\chav{1+3\Phi e^{-\beta (E_i+\mu)}+3\bar{\Phi} e^{-2\beta (E_i+\mu)}+e^{-3\beta (E_i+\mu)}}.
\end{align}
The quark condensates $\ev{\bar{q}_iq_i}$ are given by 
$$
\ev{\bar{q}_iq_i}=\ev{\bar{q}_iq_i}_{\text{vac}}+\ev{\bar{q}_iq_i}_{\text{mag}}+\ev{\bar{q}_iq_i}_{\text{med}}
$$
where
\begin{align}
 \ev{\bar{q}_iq_i}_{\text{vac}}&=-6\int_{\Lambda}\frac{d^3p}{(2\pi)^3}\frac{M_i}{E_i}\\
  \ev{\bar{q}_iq_i}_{mag}&=-\frac{3m_i|q_i|B}{2\pi^2}\Big{[} 
  \ln\Gamma(x_i)-\frac{1}{2}\ln(2\pi)+x_i\nonumber\\
  &-\frac{1}{2}(2x_i-1)\ln(x_i)\Big{]}\\
\ev{\bar{q}_iq_i}_{\text{med}}&=\frac{3(|q_i|B)^2}{2\pi}\sum_n\alpha_n\int_{-\infty}^{+\infty}
\frac{dp_z^i}{2\pi}\pc{
f_{\Phi}^+(E_i)+f_{\Phi}^-(E_i) },
\end{align}
The distribution functions $f_\Phi^{+}$ and $f_\Phi^{-}$ are
\begin{align}
 f_{\Phi}^+(E_i)&=\frac{\Phi e^{-\beta (E_i-\mu)}+2\bar{\Phi} e^{-2\beta (E_i-\mu)}+e^{-3\beta 
 (E_i-\mu)} }
 {1+3\Phi e^{-\beta (E_i-\mu)}+3\bar{\Phi} e^{-2\beta (E_i-\mu)}+e^{-3\beta (E_i-\mu)}}\\
 f_{\Phi}^-(E_i)&=\frac{\bar{\Phi} e^{-\beta (E_i+\mu)}+2\Phi e^{-2\beta (E_i+\mu)}+e^{-3\beta (E_i+\mu)}}
 {1+3\bar{\Phi} e^{-\beta (E_i+\mu)}+3\Phi e^{-2\beta (E_i+\mu)}+e^{-3\beta (E_i+\mu)}}.
\end{align}
Calculating $\frac{\partial \Omega}{\partial \phi_i}=0$, with $\phi_i=\ev{\bar{q}_uq_u}$,
$\ev{\bar{q}_dq_d}$, $\ev{\bar{q}_sq_s}$, $\Phi$ and $\bar{\Phi}$,
we obtain the gap equations
\begin{align}
\left\{
\begin{array}{l}
 M_u=m_u-2G(\Phi,\bar{\Phi})\ev{\bar{q}_uq_u}-2K\ev{\bar{q}_dq_d}\ev{\bar{q}_sq_s}\\
 M_d=m_d-2G(\Phi,\bar{\Phi})\ev{\bar{q}_dq_d}-2K\ev{\bar{q}_sq_s}\ev{\bar{q}_uq_u}\\
 M_s=m_s-2G(\Phi,\bar{\Phi})\ev{\bar{q}_sq_s}-2K\ev{\bar{q}_uq_u}\ev{\bar{q}_dq_d}\\
\end{array}	
\right.
\end{align}
\begin{align}
\frac{\partial{\cal U}}{\partial \Phi}=\frac{\partial{\cal U}}{\partial \bar{\Phi}}=0.
\end{align}
This set of coupled equations must be solved self consistently. At $\mu=0$, $\Phi=\bar{\Phi}$, and
we are left with a set of four coupled equations to solve.

\subsection{Thermodynamic quantities}

From the thermodynamic potential density $\Omega(T,\mu)$ one can derive the equations of state
which allow us to study some observables that are accessible in lattice QCD at zero chemical
potential. 
The pressure $P(T,\mu)$ is defined such that its value is zero in the vacuum 
\begin{equation}
P(T, \mu)=-\pr{\Omega(T, \mu)- \Omega(0, 0)},
\end{equation}
where $V$ is the volume of the system.

The equation of state for the entropy density $S$ is given by
\begin{equation}
S=\pc{\frac{\partial P}{\partial T}}_{\mu}
\end{equation}
and the energy density ${\cal E}$ comes from the following 
fundamental relation of thermodynamics
\begin{equation}
{\cal E}=TS+\mu\rho_B-P
\end{equation}
where the baryonic density $\rho_B$ is given by
\begin{equation}
\rho_B=-\pc{\frac{\partial \Omega}{\partial \mu}}_{T}.
\end{equation}
The interaction measure 
\begin{equation}
\Delta=\frac{{\cal E}-3P}{T^4}
\label{eq:Int}
\end{equation}
is another important quantity once it quantifies the deviation
from the equation of state of an ideal gas of massless constituents. Lattice studies
show that the interaction measure remains large even at very high temperatures,
where the Stefan-Boltzmann (SB) limit is not yet reached, and thus some interactions
must still be present. The speed of sound squared, 
\begin{equation}
v_s^2=\pc{\frac{\partial P}{\partial {\cal E}}}_V\text{,}
\end{equation}
and the specific heat, 
\begin{equation}
C_V=\pc{\frac{\partial {\cal E} }{\partial T}}_V\text{,}
\end{equation}
are important quantities that can also be calculated in lattice QCD. 
In the present study, the thermodynamic quantities are calculated at zero chemical 
potential $\mu=0$.

\section{Entanglement interaction parametrization}
\label{sec:Entanglement}

The parametrization $(\alpha_1,\alpha_2)$ of the entanglement interaction [Eq. (\ref{G})] 
was fitted in \cite{Sasaki:2011wu} with $T_0=150\unit{MeV}$, to reproduce the result of $2+1$ 
flavor LQCD at $\mu=0$ \cite{Aoki:2006we} and the results of the degenerate three-flavor LQCD 
at $\theta=\pi$ \cite{deForcrand:2010he}.
In the present work, we want to compare the EPNJL and PNJL models; therefore, we set 
$T_0=210\unit{MeV}$ in both models.
The only constraint we impose on the entanglement parametrization $(\alpha_1,\alpha_2)$
is that both the chiral and deconfinement transitions are crossovers.
To study the dependence of the order parameters $\ev{\bar{q}_iq_i}$ and $\Phi$ on the entanglement 
interaction parametrization $(\alpha_1,\alpha_2)$, we define several sets, listed in Table I, 
that we will explore in this work. These sets cover all the crossover region of the entanglement
parametrization.

\subsection{Zero magnetic field}

\begin{figure}[t]
    \includegraphics[width=1.0\linewidth,angle=0]{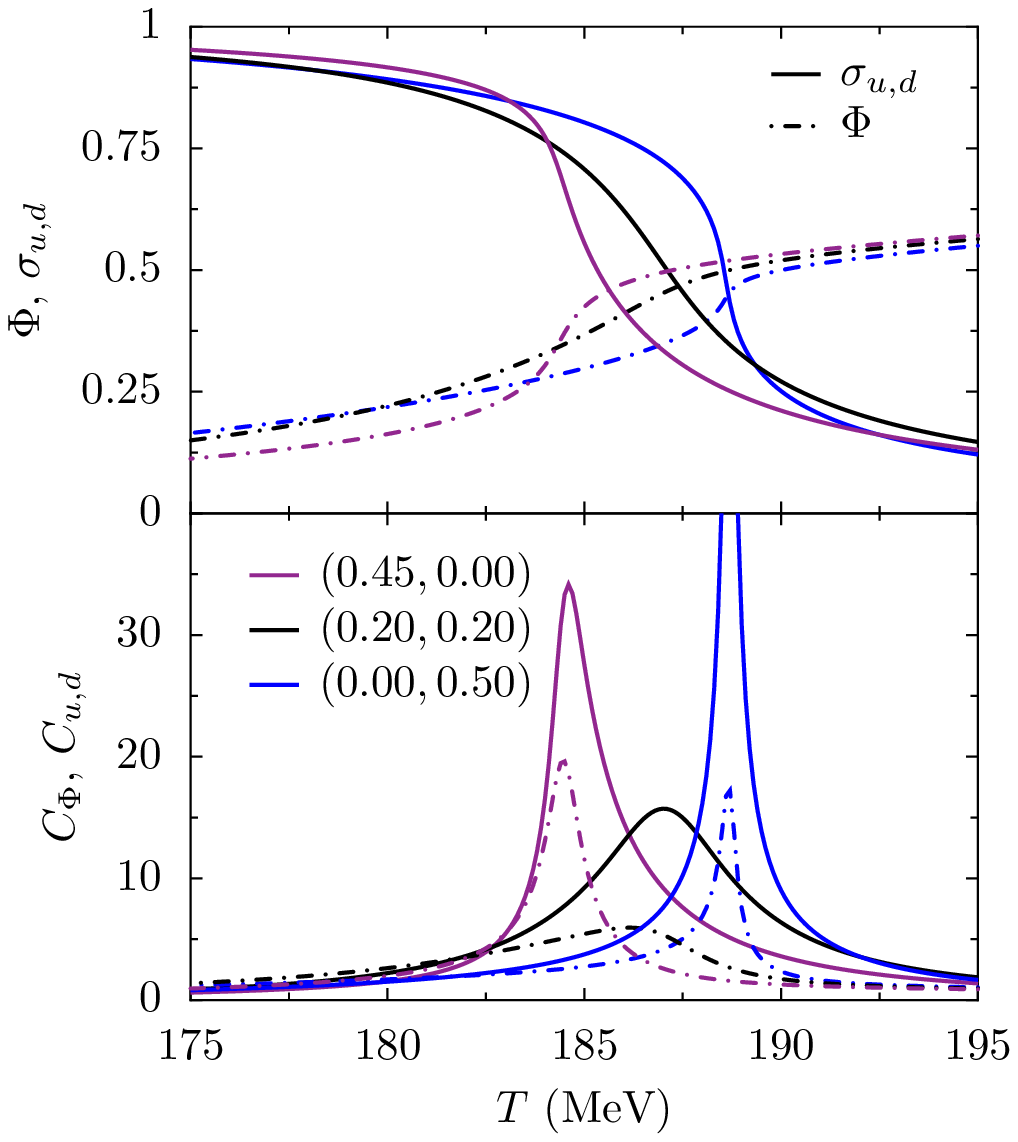}
    \caption{The vacuum normalized $u$ quark condensates $\sigma_u$ 
    (at $eB=0$ we have $\sigma_u=\sigma_d$) and the Polyakov loop $\Phi$ (top panel), and 
    the respective susceptibilities (bottom panel) for three parametrization sets
    $(\alpha_1,\alpha_2)$.}
\label{fig:condensados}
\end{figure}

The results for zero magnetic field $eB=0$ and zero baryonic chemical potential $\mu_B=0$ 
are shown in Fig. \ref{fig:condensados}, where the vacuum normalized condensates
$\sigma_i\equiv\ev{\bar{q}_iq_i}(T)/\ev{\bar{q}_iq_i}(0)$ 
(for $eB=0$, there is an exact $SU(2)$ isospin symmetry and $\ev{\bar{q}_uq_u}=\ev{\bar{q}_dq_d}$), 
the Polyakov loop $\Phi(T)$ and its susceptibilities, $C_i = -m_\pi\partial\sigma_i/\partial T$
and $C_{\Phi} = m_\pi\partial\Phi/\partial T$, are represented. The multiplication by $m_\pi$ is 
done only to ensure that the susceptibilities are dimensionless.

We have calculated the chiral and deconfinement pseudocritical 
transition temperatures, defined as the location of the peaks in $\sigma_i$ and $\Phi$
susceptibilities, at zero magnetic field. This way, the pseudocritical temperatures in the 
PNJL are $T_c^\chi=200$ and $T_c^\Phi=171\unit{MeV}$, while the results for some
parametrization sets, which cover all the crossover region, are listed in Table I.
A first conclusion from Table I is that the restoration of chiral symmetry in the EPNJL model
is influenced by the gauge fields mimicked by the Polyakov loop: 
the deconfinement transition affects the chiral transition, decreasing the interaction 
responsible for the chiral symmetry breaking and shifting the chiral symmetry restoration 
to smaller temperatures, thus, bringing both transition temperatures closer.
On the other hand, the $(0.45,0.00)$ and $(0.00,0.50)$ sets are in the limit of turning 
the crossover transitions into a first-order phase transition. This is reflected in the
susceptibilities values at the pseudocritical temperatures, being more pronounced than 
for the $(0.20,0.20)$ set.
We also notice that, even at zero magnetic field, the pseudocritical transition temperatures 
are quite sensitive to the parametrization $(\alpha_1,\alpha_2)$. For $(0.45,0.00)$ and 
$(0.00,0.50)$, they almost coincide, but for $(0.10, 0.20)$  and $(0.20, 0.10)$ we obtain 
$\Delta T_c=T^\chi_c-T^\Phi_c=3.8\unit{MeV}$ and $\Delta T_c=5.2\unit{MeV}$, respectively.
Therefore, the coincidence of the transition temperatures, the main feature of the entanglement
interaction, depends on its parametrization. We show in Fig. \ref{fig:condensados}, for 
three sets of Table I, the order parameters and their susceptibilities.

\begin{table}[t]
\begin{center}
    \begin{tabular}{|c||c|c|}
            \hline
        $(\alpha_1,\alpha_2)$    & $T^{\chi}_c [\unit{MeV}]$   & $T^\Phi_c [\unit{MeV}]$  \\
        \hline\hline
        $(0.45,0.00)$ & 184.6  & 184.5   \\
        $(0.25,0.10)$ & 186.4    & 183.6   \\
        $(0.20,0.10)$ & 187.3  &  182.1  \\
	$(0.20,0.20)$ & 187.0  &  186.2  \\
        $(0.10,0.20)$ & 188.4  &  184.6  \\
        $(0.00,0.50)$ & 188.7  &  188.7  \\
        \hline
    \end{tabular}
    \caption{\label{table:Tc} Pseudocritical temperatures for the chiral transition 
    $\left(T^\chi_c=(T^u_c+T^d_c)/2\right)$ and the deconfinement ($T^\Phi_c$)
    for several parametrization sets $(\alpha_1,\alpha_2)$ with $T_0=210$ MeV.}
\end{center}
\end{table}

\begin{figure}[t]
    \includegraphics[width=1.0\linewidth,angle=0]{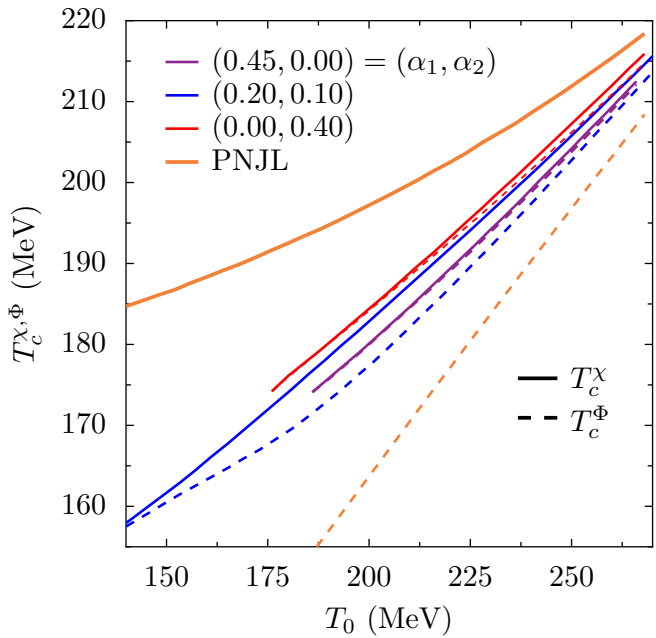}
    \caption{pseudocritical temperatures for the chiral
    $\left(T^\chi_c=(T^u_c+T^d_c)/2\right)$ and deconfinement ($T^\Phi_c$) transitions as 
    a function of $T_0$,
    for several sets $(\alpha_1,\alpha_2)$.}
\label{fig:fig2}
\end{figure}

In the following, the effect of the $T_0$ value on the EPNJL model is analyzed, in particular, 
on the transition temperatures. For that, we have calculated the pseudocritical temperatures 
of the chiral $T^\chi_c$ and the deconfinement $T^\Phi_c$ transitions as a 
function of $T_0$ for three sets of Table 1. The results are shown in Fig. \ref{fig:fig2}.
For each set $(\alpha_1,\alpha_2)$, there is a lower value of  $T_0$ ($T_0^{1st}$) that 
still gives a crossover transition for both phases. A first-order phase transition occurs 
if  $T_0<T_0^{1st}$ are used. The $T_0^{1st}$ values obtained are:
$T_0^{1st}=186$, $125$, and $176$ MeV for $(0.45,0.00)$, $(0.20,0.10)$, 
and $(0.00,0.40)$, respectively.
We see in Fig. \ref{fig:fig2} that for values of $T_0$ close to $T_0^{1st}$ both chiral and 
deconfinement transitions coincide for all sets. For higher values of $T_0$ the coincidence of 
the pseudocritical temperatures depends on the parametrization set. 
For $(0.45,0.00)$, a good coincidence is obtained for all range 
of $T_0$ but for $(0.20,0.10)$ a difference as large as $\Delta T_c\approx8$ MeV
is obtained. In Fig. \ref{fig:fig2} the result for the PNJL model
is also plotted, showing a much larger gap in $\Delta T_c$, which grows as $T_0$ decreases.

\subsection{Finite magnetic field}

Due to the different charges of the $u$ and $d$ quarks, the isospin 
symmetry is lost when an external magnetic field is applied to the system.
Thus, when $eB\neq0$ we get $\ev{\bar{q}_uq_u}(B,T)\neq\ev{\bar{q}_dq_d}(B,T)$
and the chiral transition for $u$ and $d$ quarks do not coincide anymore. 
We are going to study how the magnetic field affects the transition temperatures 
and how it depends on the entanglement interaction parametrization.

\begin{figure}[t]
    \includegraphics[width=1.0\linewidth,angle=0]{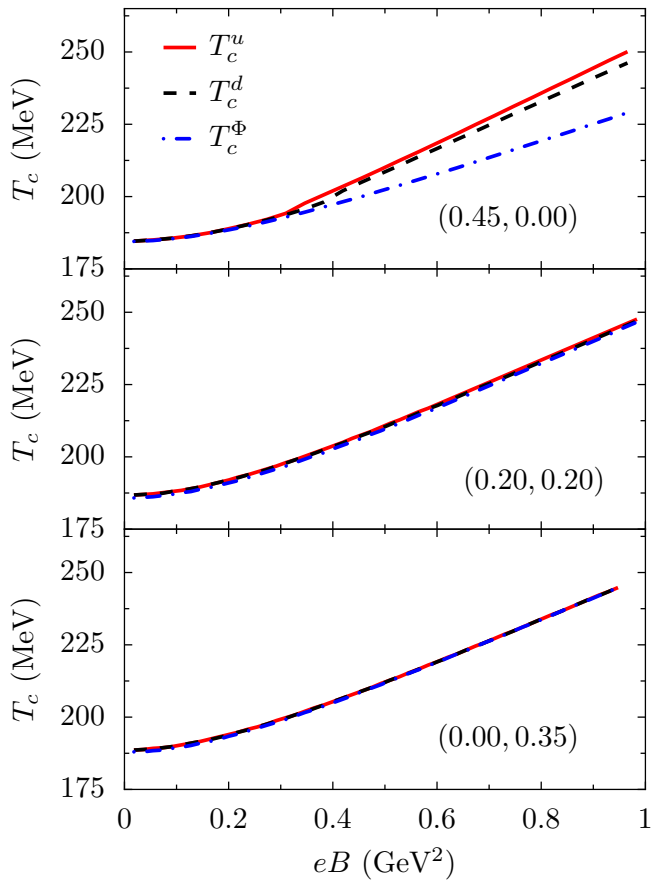}
    \caption{Transition temperatures as a function of the magnetic 
    field for three sets: 
    $(0.00,0.35)$ (bottom panel), $(0.20,0.20)$ (middle panel) and 
    $(0.45,0.00)$ (top panel).}
\label{fig:fig3}
\end{figure}

The transition temperatures as a function of the magnetic field $eB$, for 
$T_0=210\unit{MeV}$ (hereafter we use $T_0=210\unit{MeV}$ in both models), are shown
in Fig. \ref{fig:fig3}, for three sets: $(0.45,0.00)$, $(0.20,0.20)$, and
$(0.00,0.35)$. 
The transition temperatures coincide for $(0.20,0.20)$ and $(0.00,0.35)$, even
with a finite magnetic field. In the last set, for $eB>0.91 \unit{GeV}^2$, we obtain
a first-order phase transition, and for lower values the coincidence in the 
transition temperatures is perfect. For $(0.45, 0.00)$, unlike the other sets,
the magnetic field, at $eB\approx 0.3 \unit{GeV}^2$, breaks the coincidence of the 
chiral and deconfinement transitions, and the deconfinement temperature 
is less affected than the chiral transition temperature, even though the magnetic 
field enhances both the condensates and Polyakov loop.

The region between the chiral and deconfinement transitions can be called the constituent
quark phase (CQP) \cite{Kouno:1988bi,Cleymans:1986cq}, 
where the deconfinement already occurred but the chiral symmetry remains broken.  

As a result of the charge difference between $u$ and $d$ quarks, we obtain a higher 
transition temperature for the $u$ than the $d$ quark, and this difference grows as 
the magnetic field increases. 
This pattern was also found in the context of the instanton-liquid model, modified by the
Harrington-Shepard caloron solution at finite $T$ in the chiral limit \cite{Nam:2011vn}, 
or in the Sakai-Sugimoto model \cite{Callebaut:2013ria}.

In the present model, the chiral transition temperature
increases with the magnetic field just as in several effective models 
\cite{Mizher:2010zb,Miransky:2002rp,Andersen:2012jf,Skokov:2011ib},
and some LQCD studies \cite{D'Elia:2010nq}.
This behavior is due to magnetic catalysis, that is, the magnetic field enhances the condensate and 
this effect explains the increase of the transition temperature with the magnetic field.
The strength of the magnetic catalysis depends on the flavor due to the charge difference.

Recent LQCD results show the inverse mechanism (inverse magnetic catalysis) near the transition
temperatures, that is, the condensate shows a nonmonotonic behavior near the transition 
temperatures, decreasing with $eB$ near the transition temperature. 
Thus, a decreasing dependence of the chiral transition temperature on magnetic field was
obtained in LQCD \cite{Bali:2011qj}. 
In \cite{Ilgenfritz:2013ara} new lattice QCD calculations report a rise of the Polyakov 
loop with $eB$ at the pseudocritical temperature and $eB\lesssim0.8$ GeV$^2$ indicating an 
inverse magnetic catalysis. 
However, at sufficiently strong magnetic field strength the magnetic catalysis is seen
in agreement with almost all effective models that predict magnetic catalysis at 
any temperature and magnetic field strength.

\begin{figure}[t]
    \includegraphics[width=1.0\linewidth,angle=0]{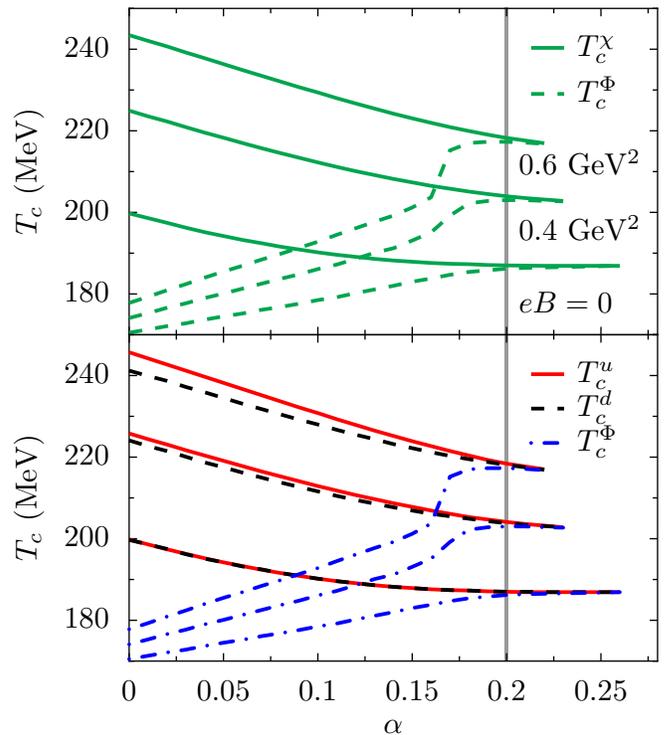}
    \caption{Transition temperatures $T_c^i$ (bottom panel), $T^\chi_c$ 
    and $T^\Phi_c$ (top panel) 
    as a function of $(\alpha,\alpha)$ for
    $eB=0,0.4$ and $0.6\unit{GeV}^2$. The grey line is the region
    plotted in the middle panel of the Fig. \ref{fig:fig3}}
\label{fig:fig4}
\end{figure}

In the following, the dependence of the pseudocritical temperatures on the entanglement
parametrization $(\alpha_1,\alpha_2)$ is calculated.
First, we set $\alpha_1=\alpha_2=\alpha$, and calculate the transition temperatures as a function of
$\alpha$ for three magnetic field intensities: $eB=0,0.4$ and $0.6\unit{GeV}^2$. The results 
are shown in Fig. \ref{fig:fig4}. As $\alpha$ increases, the 
deconfinement temperature increases and the chiral transition decreases. At some critical value
of $\alpha$, for $eB=0.4$ and $0.6\unit{GeV}^2$, the gap between both critical temperatures 
decreases abruptly before the first-order phase transition sets in. 
The gray line of Fig. \ref{fig:fig4} is the region plotted in the middle panel of the 
Fig. \ref{fig:fig3}.

In \cite{Gatto:2010pt}, the effect of varying the entanglement parametrization was already studied 
by using the SU(2) PNJL model with and without eight-quark interaction term.  
As in the present work, the existence of a value of $\alpha$ was found, where the crossover 
is replaced by a first-order phase transition, which depends on the magnetic field strength.
Fig. \ref{fig:fig4} also shows that this value of $\alpha$ ($\alpha^{\text{1st}}$) 
depends on $eB$ getting smaller with increasing $eB$. 
For $\alpha=0$ the EPNJL model reduces to the PNJL model, because $G(\Phi)$ reduces to $G$.
We see that the EPNJL model always gets a smaller gap in $\Delta T_c=T^\chi_c-T^\Phi_c$ 
than the PNJL model, for any magnetic field strength. 
The ratio $G(\Phi)/G$ is always equal or smaller than one, which means that the
parameter responsible for the chiral symmetry breaking in the (P)NJL model
is always larger than the one in the EPNJL model. 

The coincidence or not of the phase transition temperatures depends on the parametrization chosen.
Therefore, the existence or not of the CQP phase depends on the entanglement parametrization. 
With a particular choice of $(\alpha_1,\alpha_2)$, the CQP phase can be included or removed from
the phase diagram.

\begin{figure*}[t!]
    \includegraphics[width=0.46\linewidth,angle=0]{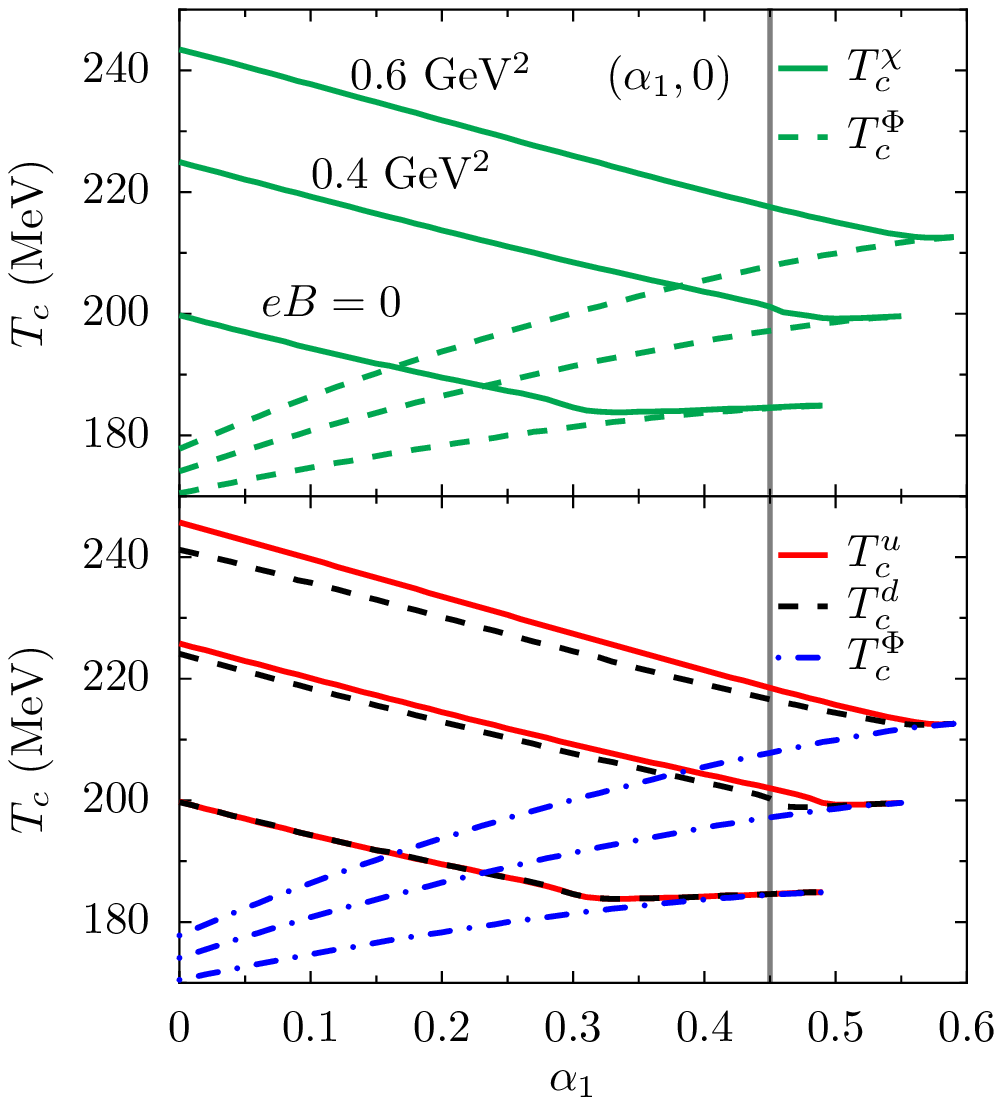}
    \includegraphics[width=0.45\linewidth,angle=0]{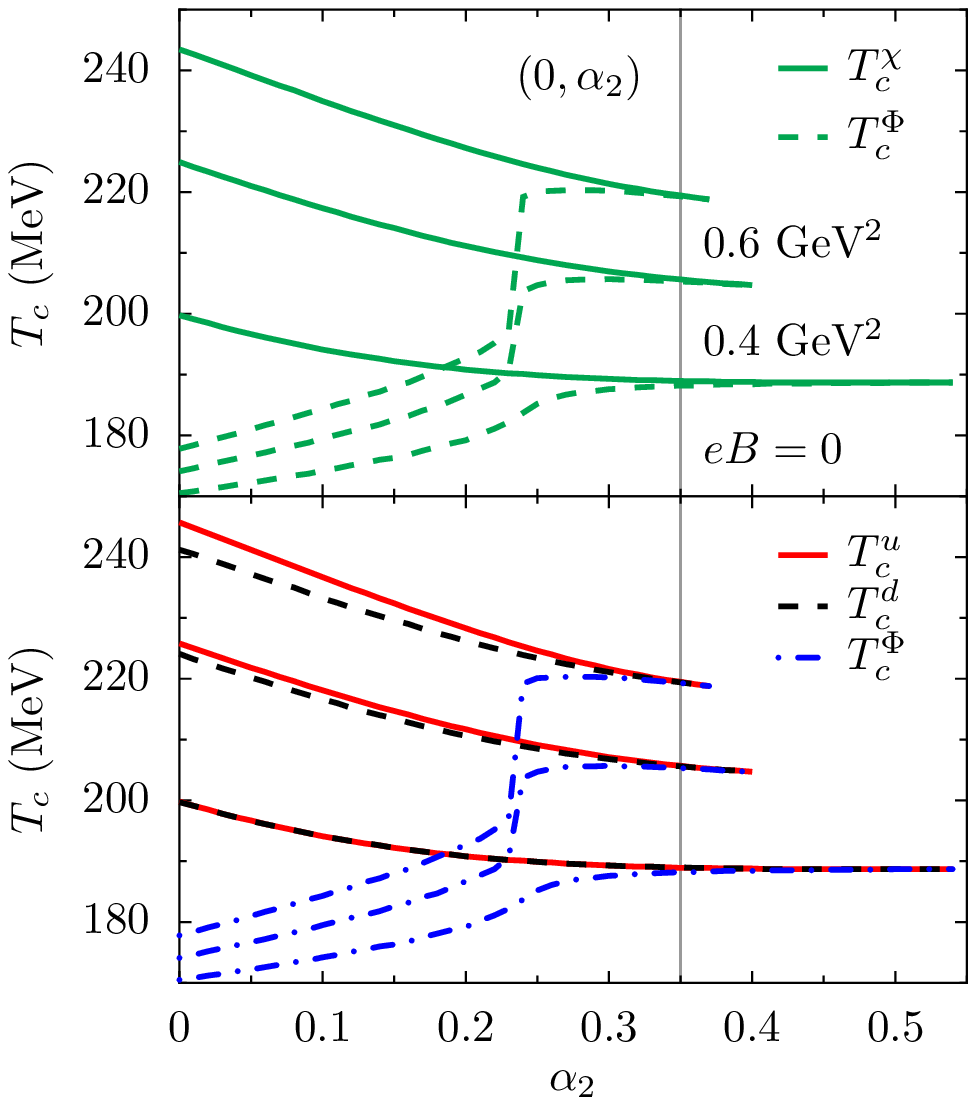}
    \caption{Transition temperatures $T^\chi_c$ and $T^\Phi_c$ (top panel) and
    $T_c^i$ with $i=u,d,\Phi$ (bottom panel) as a function of $(0,\alpha_2)$ 
    (right) and $(\alpha_1,0)$ (left) for $eB=0,0.4$ and $0.6\unit{GeV}^2$. 
    The gray lines are the regions plotted in the top and bottom panels of the 
    Fig. \ref{fig:fig3}.}
\label{fig:fig5}
\end{figure*}

Now, we set $\alpha_1=0$ or $\alpha_2=0$, and calculate the transition temperatures as a 
function of $(0,\alpha_2)$ and $(\alpha_1,0)$, respectively. With $\alpha_1=0$ or $\alpha_2=0$, 
we are [by Eq. (\ref{G})] choosing the functional form of the entanglement interaction as
$G(\Phi)\propto\alpha_2\Phi^3$ or $G(\Phi)\propto\alpha_1\Phi^2$, respectively.
The results are in in Fig. \ref{fig:fig5} and show the following two main differences:\\
(a) for $(\alpha_1,0)$, (Fig. \ref{fig:fig5}, left panel), the $\alpha_1^{\text{1st}}$ 
increases with increasing $eB$, making it possible that for weak magnetic fields 
the crossover transition turns into first-order phase transition. Nevertheless, if
for $eB=0$ a crossover is obtained, it will always remain a crossover even when $eB$
is increased. However, some parametrizations of $(\alpha_1,0)$ allow a first-order phase 
transition for low $eB$, while a crossover is obtained for higher values of $eB$. 
For $(0,\alpha_2)$ (right panel of Fig. \ref{fig:fig5}), $\alpha_2^{\text{1st}}$
has the opposite behavior so it is possible to have a set of $(0,\alpha_2)$ values where for
$eB=0$ a crossover is obtained but a first-order phase transition exists when $eB$ increases.
This behavior is qualitatively similar to the one found for $(\alpha,\alpha)$ shown in Fig. 
\ref{fig:fig4}.\\
(b) for a fixed $eB$, the gap $\Delta T_c$ decreases as $\alpha_1$ or $\alpha_2$ increases, but
for $(\alpha_1,0)$ (left panel of Fig. \ref{fig:fig5}) the $T^\Phi$ always increases without any 
bump and for high values of $\alpha_1$, closer to the first-order phase transition, it is the
$T^\chi$ that follows the $T^\Phi$, contrarily to what happens in the case $(0,\alpha_2)$.  

The gray lines in both panels of Fig. \ref{fig:fig5} are the parametrizations explored in Fig. 
\ref{fig:fig3}. The behavior of $(0.45, 0.00)$ and $(0.00, 0.35)$ of Fig. \ref{fig:fig3} 
become now clear: for $(0.45, 0.00)$ (upper panel of Fig. \ref{fig:fig3}), at low $eB$, we are
close to the first-order phase transition, with increasing $eB$, the $\alpha_{\text{1st}}$
increases and we are moving into the crossover region where there is a $\Delta T_c$ gap; for
$(0.00, 0.35)$, we are close to $\alpha_{\text{1st}}$ and there is no $\Delta T_c$ gap,
with increasing $eB$, the $\alpha_{\text{1st}}$ decreases, and for $eB>0.91 \unit{GeV}^2$, when 
$\alpha_{\text{1st}}<0.35$, we get first-order phase transitions.

Finally a word on the pseudocritical temperature for the chiral transition corresponding to the
heavier $s$ quark. Due to the $s$ quark larger mass, the $s$ sector shows a much weaker
transition than the $u$ and $d$ sectors [$C_s(T)<<C_{u,d}(T)$], being also the respective
pseudocritical temperature $T_c^s>T_c^{u,d}$ only slightly affected by the increase of $eB$ 
within the range considered in the present work.

\section{Thermodynamics}

In the following, we are going to study the behavior of several thermodynamical quantities in 
the presence of an external magnetic field $eB$ at zero chemical potential $\mu=0$, that is, 
in the $T-eB$ plane. The dependence of these properties on the parametrization of the 
entanglement interaction will be also discussed.

\begin{figure*}[t]
    \includegraphics[width=0.355\linewidth,angle=0]{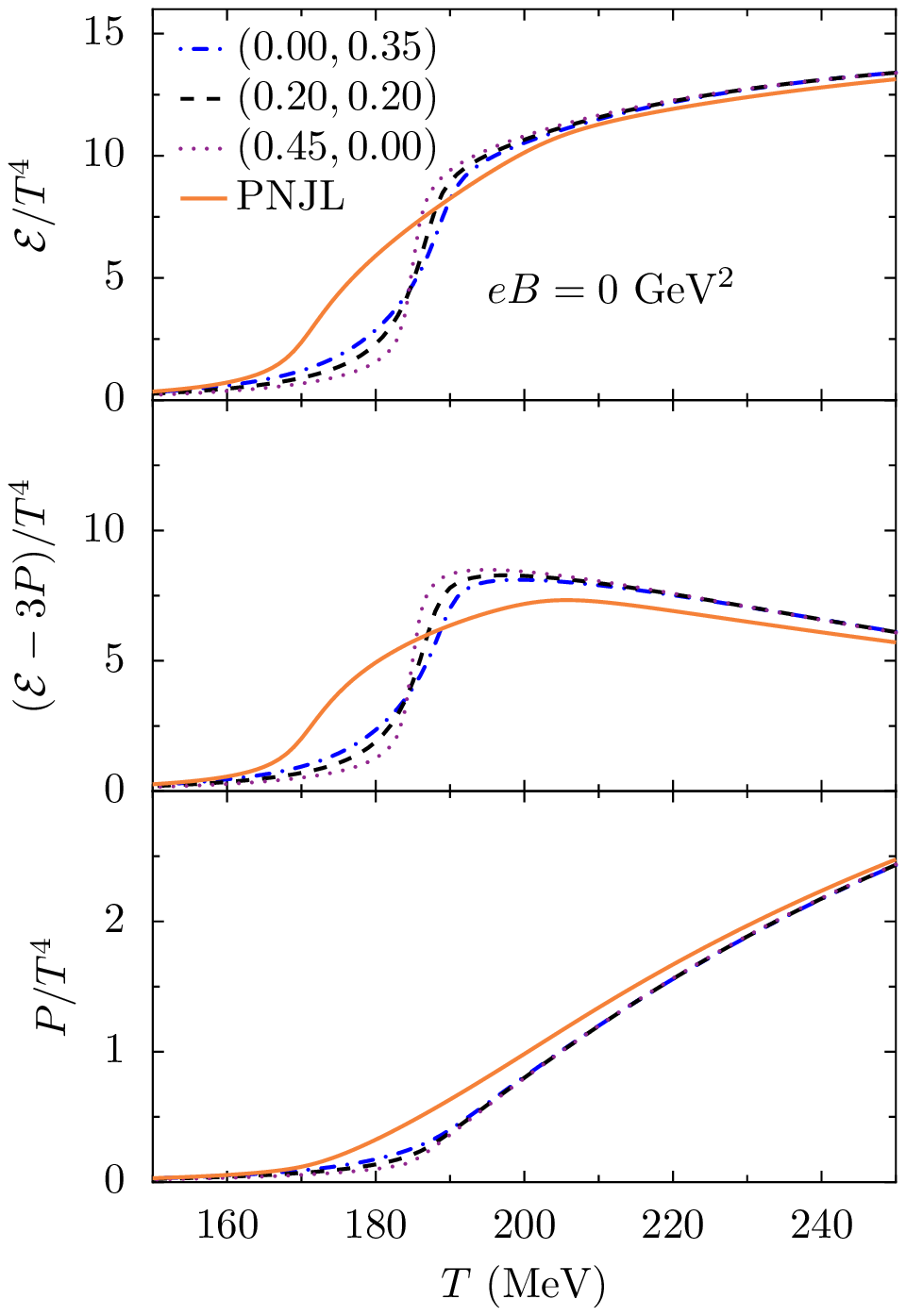}
    \includegraphics[width=0.31\linewidth,angle=0]{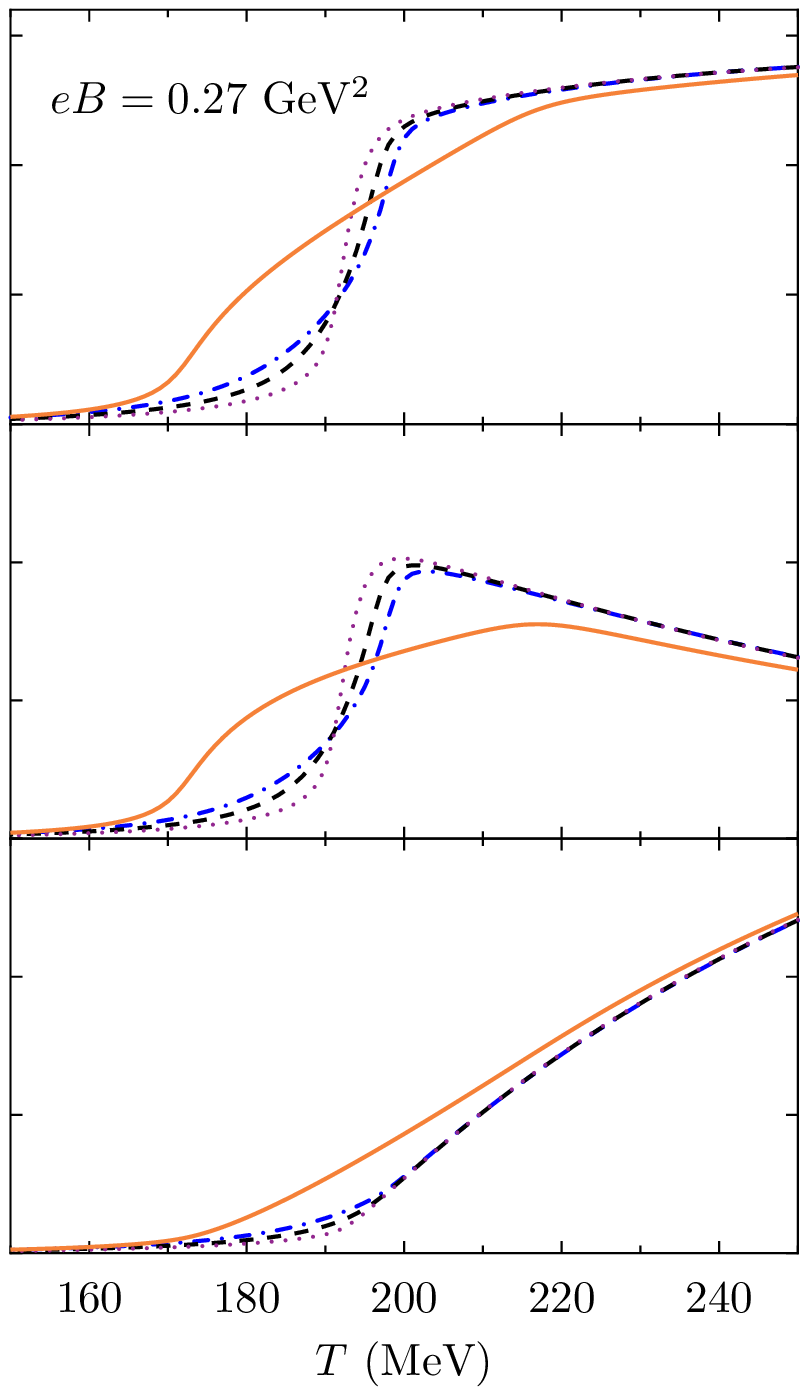}
    \includegraphics[width=0.31\linewidth,angle=0]{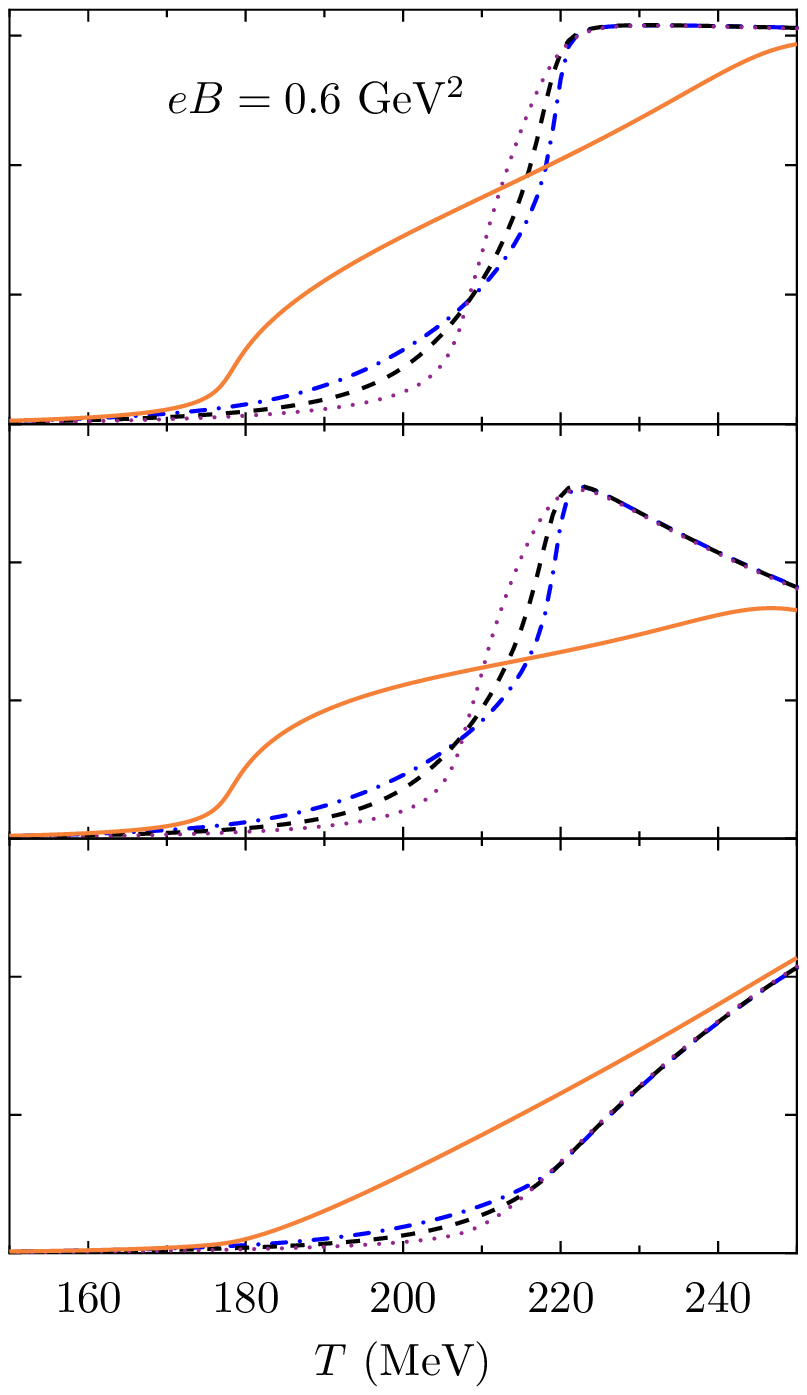}
		\caption{The scaled energy density ${\cal E}/T^4$, the interaction measure 
		$\Delta(T)=({\cal E}-3P)/T^4$, and the scaled pressure $P/T^4$ as a function
		of temperature $T$, for $eB=0, 0.27$ and $0.6\unit{GeV}^2$ in (E)PNJL models.}
\label{fig:press_energia}
\end{figure*}

In Fig. \ref{fig:press_energia}, we plot the scaled pressure $P/T^4$, the scaled energy
density ${\cal E}/T^4$, and the interaction measured $\Delta$ [Eq. (\ref{eq:Int})] 
as a function of temperature for $eB=0$, so we can compare compare the EPNJL parametrizations with 
the PNJL model \cite{costa}, for $eB\approx0.27\unit{GeV}^2$, being this value an estimation 
of the maximal magnetic field strength for the LHC \cite{Skokov:2009qp} and $0.6\unit{GeV}^2$, 
an already high magnetic field. 

Since the transition to the high temperature phase is a rapid crossover rather than a 
phase transition, the pressure, the energy density and thus the interaction measure
are continuous functions of the temperature. We observe a similar behavior in the three
curves for the EPNJL model for the different scenarios: 
a sharp increase in the vicinity of the transition temperature 
and then a tendency to saturate at the corresponding ideal gas limit. 
The sharp increase in the PNJL model occurs at lower temperatures than the EPNJL due to 
the difference in the deconfinement transition temperature given by both models,
$T_c^\Phi=171\unit{MeV}$ in the PNJL and $T_c^\Phi=182-189\unit{MeV}$
in the EPNJL. The energy density rises sharply above the transition temperature in the EPNJL. 
At $eB=0.6\unit{GeV}^2$, in the PNJL model, the energy density shows two bumps, corresponding to 
deconfinement and chiral transitions, that at $eB=0.6\unit{GeV}^2$
are $T_c^\Phi=178$ and $T_c^\chi=244\unit{MeV}$, respectively. 

\begin{figure*}[t]
    \includegraphics[height=0.35\linewidth,angle=0]{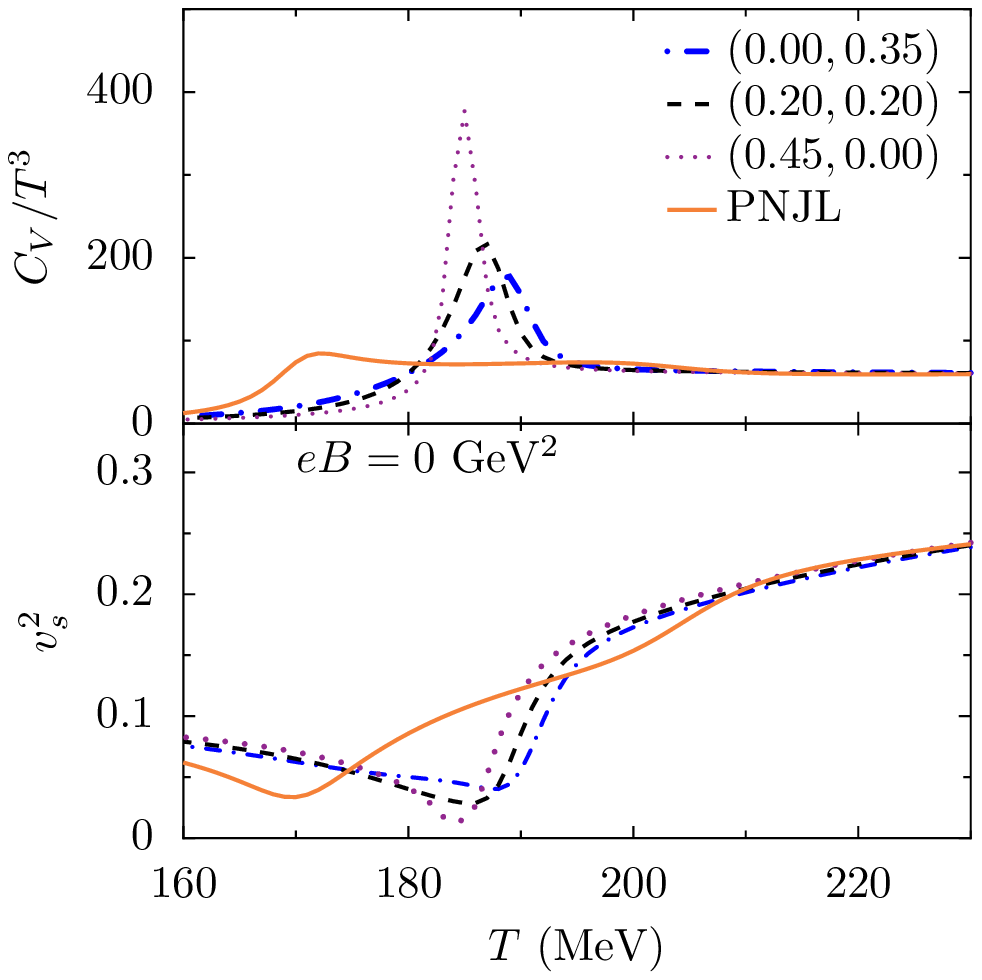}
    \includegraphics[height=0.35\linewidth,angle=0]{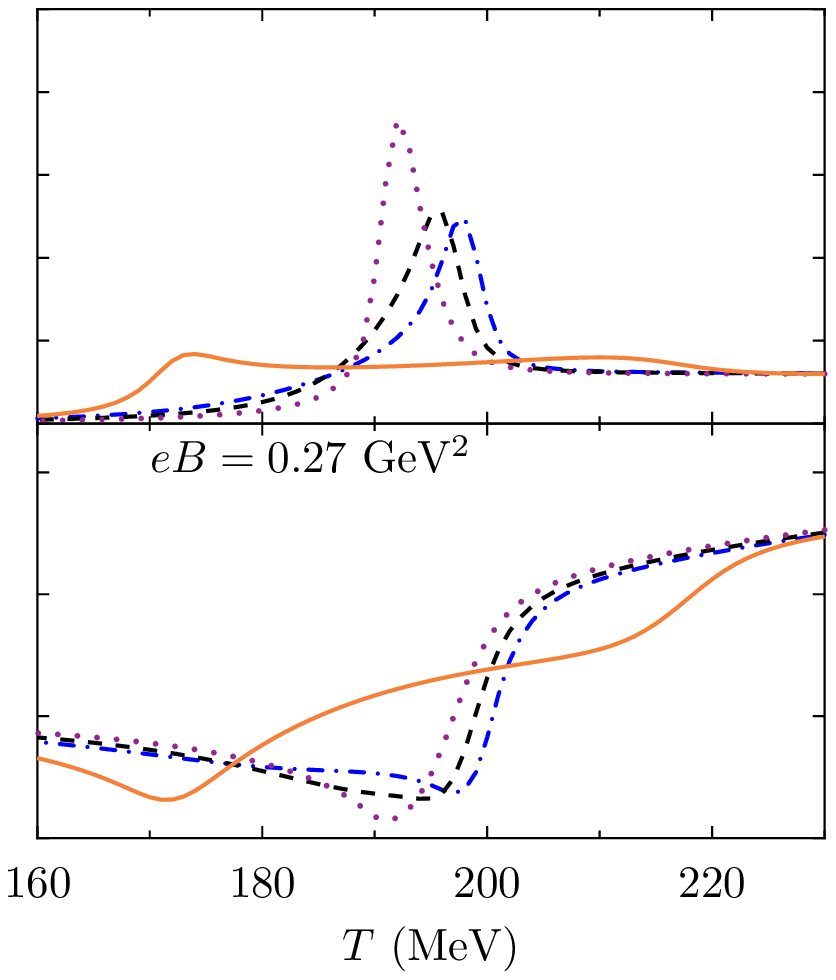}
    \includegraphics[height=0.35\linewidth,angle=0]{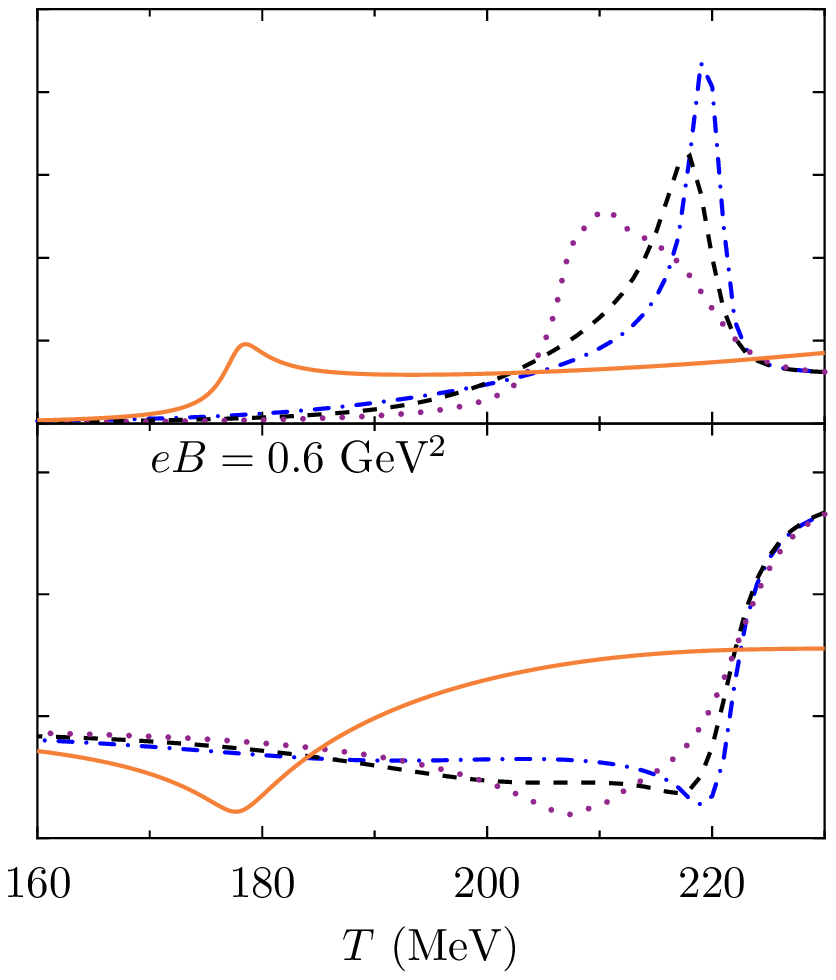}
		\caption{The scaled specific heat $C_V/T^3$, and speed of sound squared $v_s^2$ as a 
		function of temperature $T$, for $eB=0, 0.27$ and $0.6\unit{GeV}^2$ in (E)PNJL models.}
\label{fig:cv_ss}
\end{figure*}

Fig. \ref{fig:cv_ss} shows the scaled specific heat $C_V/T^3$ and the 
speed of sound squared $v_s^2$ as a function of the temperature, for 
$eB=0, 0.27$ and $0.6\unit{GeV}^2$. In both models, at high temperatures,
a common limit is obtained the two observables. This was expected due to the same number
of degrees of freedom in both models. The specific heat increases strongly
near the deconfinement temperature and, at $eB=0$, it is much higher in the 
EPNJL model. However, as the magnetic field increases, the $C_V$
in the PNJL model increases to values near the ones in the EPNJL.
Once more, we see that the PNJL model shows two peaks in $C_V$ at any $eB$, caused 
by the distinct chiral and deconfinement transitions. The first peak is due
to the deconfinement and the second to the chiral transition.
The speed of sound squared $v_s^2$ passes through a local minimum around the deconfinement 
temperature and then reaches the limit of $1/3$ (SB limit) at high temperature. 
This minimum signals a fast change in the masses of quarks in both EPNJL and PNJL models. 
The pattern of local minim, shown by $v_s^2$ as a function of the magnetic field, is 
related to the temperatures at which both phase transitions occur, as in the case of the 
peaks of $C_V/T^3$.

For the EPNJL, it is interesting to look at each parametrization. For $(0.45,0.00)$, we know
from the top panel of Fig. \ref{fig:fig3} that $T_c^\Phi$ and $T_c^\chi$ coincide, at low $eB$
and not at high $eB$. This is also reflected in the quantities $C_V/T^3$ and $v_s^2$:
at $eB=0$, it has the maximum $C_V/T^3$ from all parametrizations, but it decreases 
as we increase $eB$; at $0.6\unit{GeV}^2$, aside from having the lowest value, 
it has the broadest peak, signaling the the increasing $\Delta T_c$ gap with $eB$. 
The $(0.00,0.35)$ parametrization has the lowest $C_V/T^3$ peak at $eB=0$, 
but the highest at $eB=0.6\unit{GeV}^2$, showing that the parametrization keep the 
$\Delta T_c$ gap close to zero at any magnetic field strength (see middle panel of 
Fig. \ref{fig:fig3}), and with increasing $eB$ the first-order phase transitions become closer.
At last, for the $(0.20,0.20)$ parametrization, the maximum value of $C_V/T^3$ increases 
slightly with $eB$. Looking at Fig.
\ref{fig:fig4}, we see that at $eB=0.6\unit{GeV}^2$ we have $\alpha_{\text{1st}}>0.20$; that is, 
we are in the crossover region for magnetic fields up to $0.6\unit{GeV}^2$.

\section{Conclusions}

In this work we have studied the three-flavor quark matter under the influence of an 
external magnetic field using the EPNJL model. The pseudocritical temperatures have 
been calculated as a function of the magnetic field strength  and the range of 
possible parametrizations of the entanglement interaction analyzed. 
The main result obtained is the conclusion that the coincidence or not of the 
deconfinement and chiral pseudocritical temperatures, also including the effect 
of the magnetic field, depends on the parametrization chosen. 
The PNJL model predicts different critical temperatures for both phase transitions and
their difference increases as the magnetic field strength grows. Within the PNJL model,
the temperature of the deconfinement transition is almost insensitive to the magnetic field 
when compared with the chiral transition temperature. 

In a LQCD study \cite{D'Elia:2010nq}, it was found that the transition temperature
increases slightly with the magnetic field (this study shows no inverse magnetic catalysis) 
and no evidence for a disentanglement of both phase transitions  was found, 
at least for magnetic fields up to $\sim0.36$ GeV$^2$. 
It was also observed that the transition becomes a sharper crossover and a first-order 
phase transition might appear. 
Therefore, as pointed out in \cite{Gatto:2010pt}, the parametrization of the entanglement 
interaction can be used to reproduce this behavior. 
However, in order to reproduce the inverse magnetic catalysis, as some recent LQCD results 
\cite{Bali:2011qj} show, the model must be modified once the entanglement interactions between 
the Polyakov loop and the chiral condensate are not able to describe the inverse magnetic
catalysis. A way to modify the model was proposed in \cite{Ferreira:2013tba}, where a magnetic field
dependent $T_0(eB)$ allows us to mimic the reaction of the gluon sector to the presence of an external
magnetic field in order to reproduce the correct behavior of transition temperatures given by
lattice QCD. Nevertheless, this same mechanism also can give rise to a first-order phase 
transition at quite low magnetic fields.

Finally, as expected, the entanglement interaction also affects the thermodynamic properties.
In particular, we have shown that the dependence on temperature of
the heat capacity and sound velocity are sensitive to the entanglement interaction.
Both quantities reflect the smaller or larger coincidence between chiral and deconfinement
transitions, and the proximity of a first-order phase transition.

\vspace{1cm}
{\bf Ackowledgements}: 
This work was partially supported by Projects Nos. PTDC/FIS/ 113292/2009 and
CERN/FP/123620/2011 developed under the initiative QREN financed by the UE/FEDER 
through the program COMPETE$--$``Programa Operacional Factores de Competitividade''$--$
and by Grant No. SFRH/BD/51717/2011.


\end{document}